\documentclass[aps,prb,twocolumn,superscriptaddress,fleqn]{revtex4}

\usepackage[english]{babel}
\usepackage{amsmath}
\usepackage{amssymb}
\usepackage{graphicx} 
\usepackage{siunitx}
\usepackage[usenames, dvipsnames]{color}
\usepackage{bm}

\begin{document}
\title{Dual-energy electron beams from a compact laser-driven accelerator}

\author{J. Wenz$^{*}$}
\address{Ludwig-Maximilians-Universit\"at M\"unchen, Am Coulombwall 1, 85748 Garching, Germany}
\address{Max Planck Institut f\"ur Quantenoptik, Hans-Kopfermann-Str.\ 1, Garching 85748, Germany}
\author{A. D\"opp$^{*,\dagger}$}
\address{Ludwig-Maximilians-Universit\"at M\"unchen, Am Coulombwall 1, 85748 Garching, Germany}
\address{Max Planck Institut f\"ur Quantenoptik, Hans-Kopfermann-Str.\ 1, Garching 85748, Germany}
\author{K. Khrennikov$^{*}$}
\address{Ludwig-Maximilians-Universit\"at M\"unchen, Am Coulombwall 1, 85748 Garching, Germany}
\address{Max Planck Institut f\"ur Quantenoptik, Hans-Kopfermann-Str.\ 1, Garching 85748, Germany}
\author{S. Schindler}
\address{Ludwig-Maximilians-Universit\"at M\"unchen, Am Coulombwall 1, 85748 Garching, Germany}
\address{Max Planck Institut f\"ur Quantenoptik, Hans-Kopfermann-Str.\ 1, Garching 85748, Germany}
\author{M. Gilljohann}
\address{Ludwig-Maximilians-Universit\"at M\"unchen, Am Coulombwall 1, 85748 Garching, Germany}
\address{Max Planck Institut f\"ur Quantenoptik, Hans-Kopfermann-Str.\ 1, Garching 85748, Germany}
\author{ H. Ding}
\address{Ludwig-Maximilians-Universit\"at M\"unchen, Am Coulombwall 1, 85748 Garching, Germany}
\address{Max Planck Institut f\"ur Quantenoptik, Hans-Kopfermann-Str.\ 1, Garching 85748, Germany}
\author{J. G\"otzfried}
\address{Ludwig-Maximilians-Universit\"at M\"unchen, Am Coulombwall 1, 85748 Garching, Germany}
\author{A. Buck}
\address{Ludwig-Maximilians-Universit\"at M\"unchen, Am Coulombwall 1, 85748 Garching, Germany}
\address{Max Planck Institut f\"ur Quantenoptik, Hans-Kopfermann-Str.\ 1, Garching 85748, Germany}
\author{J. Xu}
\address{Max Planck Institut f\"ur Quantenoptik, Hans-Kopfermann-Str.\ 1, Garching 85748, Germany}
\address{State Key Laboratory of High Field Laser Physics, Shanghai Institute of Optics and Fine Mechanics, Chinese Academy of Sciences, P. O. Box 800-211, Shanghai 201800, China.}
\author{M. Heigoldt}
\address{Ludwig-Maximilians-Universit\"at M\"unchen, Am Coulombwall 1, 85748 Garching, Germany}
\address{Max Planck Institut f\"ur Quantenoptik, Hans-Kopfermann-Str.\ 1, Garching 85748, Germany}
\author{W. Helml}
\address{Ludwig-Maximilians-Universit\"at M\"unchen, Am Coulombwall 1, 85748 Garching, Germany}
\address{Technische Universit\"at Dortmund, Maria-Goeppert-Mayer-Str. 2, 44227 Dortmund, Germany}
\address{Technische Universit\"at M\"unchen, James-Franck-Str.\ 1, 85748 Garching, Germany}
\author{L. Veisz}
\address{Max Planck Institut f\"ur Quantenoptik, Hans-Kopfermann-Str.\ 1, Garching 85748, Germany}
\address{Department of Physics, Umea University, SE-901 87 Umea, Sweden}
\author{ S. Karsch$^{\dagger}$}
\address{Ludwig-Maximilians-Universit\"at M\"unchen, Am Coulombwall 1, 85748 Garching, Germany}
\address{Max Planck Institut f\"ur Quantenoptik, Hans-Kopfermann-Str.\ 1, Garching 85748, Germany}

\begin{abstract}
Ultrafast pump-probe experiments open the possibility to track fundamental material behaviour like changes in its electronic configuration in real time. To date, most of these experiments are performed using an electron or a high-energy photon beam, which is synchronized to an infrared laser pulse. Entirely new opportunities can be explored if not only a single, but multiple synchronized, ultra-short, high-energy beams are used. However, this requires advanced radiation sources that are capable of producing dual-energy electron beams, for example. Here, we demonstrate simultaneous generation of twin-electron beams from a single compact laser wakefield accelerator. The energy of each beam can be individually adjusted over a wide range and our analysis shows that the bunch lengths and their delay inherently amount to femtoseconds. Our proof-of-concept results demonstrate an elegant way to perform multi-beam experiments in future on a laboratory scale.
\end{abstract}

\maketitle


Understanding the dynamics of materials on the time scale of electronic, atomic and molecular motion is one of the grand challenges of contemporary physics, chemistry and biology.\cite{Krausz:2009hz} A particularly useful tool to study these phenomena are pump-probe experiments, where a process is triggered using a pump pulse and its temporal evolution is subsequently examined using a probe pulse. Importantly, the properties of the radiation pulses dictate which type of systems can be studied with this method. Commonly available short-pulse infrared lasers are mostly used as pump to excite or manipulate weakly-bound electronic\cite{Kupitz:2014ed} and magnetic\cite{Beaurepaire:1996vo} states, to ionize\cite{Pertot:2017fw} or to heat a target\cite{Rousse:2001tq}. The induced dynamics of the system are probed with a short electron or photon beam. For instance, electron\cite{Sciaini:2011hi} or X-ray\cite{Cavalleri:2006da} diffraction are sensitive to atomic arrangement, while X-ray absorption spectroscopy\cite{Bressler:2004hz} is a particularly useful tool to study complex systems, because materials exhibit well-distinguishable transitions in the X-ray regime, i.e. element selectivity\cite{Chergui:2017jt}. 

In the case of processes governed by atomic motion, the vibrational period ($\SI[parse-numbers=false]{\sim 100}{\femto\second}$) needs to be resolved\cite{Rousse:2001uw}. So far, the required femtosecond X-ray pulses for pump-probe experiments could only be provided at accelerator-based lightsources\cite{Bostedt:2016it}, using either femto-slicing beamlines \cite{Schoenlein:2000wx} or free-electron lasers (FELs)\cite{McNeil:2010dp,Pellegrini:2017kf}. In the near future, laser-driven accelerators\cite{Esarey:2009ks} can serve as complementary or alternative femtosecond radiation sources. Their ultrashort\cite{Lundh:2011js} MeV-to-GeV-scale\cite{Malka:2002eu,Leemans:2006ux} electron beams are already being used to provide femtosecond photon beams in the THz\cite{Leemans:2003ij}, ultraviolet\cite{Fuchs:2009da}, X-ray\cite{Khrennikov:2015gxa,Dopp:2017dza} and $\gamma$-ray\cite{Yan:2017dn} regimes. While laser-driven X-ray sources were initially limited to performing basic radiography\cite{Fourmaux:2011cs,Kneip:2011cx}, recent experiments have started to take advantage of the sources' temporal resolution in pump-probe studies of warm dense matter\cite{Mahieu:2018bx} and laser-driven shock waves\cite{Wood:2018gs}.

An entirely new class of experiments becomes available when short-wavelength pulses are used for both pumping and probing\cite{Allaria:2013jp,Bencivenga:2015fq,Ferrari:2016jw}. In this dual-color operation it is for instance possible to combine direct stimulation of core state transitions with the sensitivity of the above-mentioned X-ray probing techniques. Motivated by this perspective, the FEL community has been actively developing a number of different schemes for dual-color operation\cite{Hemsing:2014dq}, ranging from multiple seed pulses\cite{Ferrari:2015df}, over staggered undulator magnets \cite{Inubushi:2013gf} to dual-energy electron bunches\cite{Ratner:2015gpa}. The latter have the great advantage that they require no modifications to the undulator beamline and the pulse delay can be adjusted with magnetic chicanes\cite{Ronsivalle:2014ks}. But even in this case the separation of both electron beam energies is at the percent level. Accordingly, the X-ray energy separation is also at the percent-level for this approach, while up to 30 percent can be reached using staggered undulators. Furthermore, access to free-electron lasers is restricted to a few experimental groups per year, and the demand of round-the-clock operation severely limits their flexibility in terms of setup changes between beam times. Thus, it is very difficult to explore new or unconventional experimental configurations such as dual-color operation. A complementary compact source of widely tunable, ultrashort, dual-energy electron beams, is therefore highly desirable to pursue high-energy pump-probe experiments of molecular or atomic systems.

Here we demonstrate the generation of such twin-beams from a laser-wakefield accelerator (LWFA) driven by a 100-TW-class Ti:Sa laser. The energy of the two beams can be adjusted over a wide energy range and our analysis shows that both duration and temporal separation of the beams is of the order of \SI{10}{\femto\second}. Given the large energy difference, the beams should be suitable for timing adjustments in magnetic chicanes. We also discuss the performance of dual-color photon sources derived from these twin-beams and present a first implementation in an all-optical Compton source. 

\section*{Comparison of monoenergetic electron beam sources}

 Over the past decade, a variety of techniques has been introduced to generate narrow bandwidth electron beams using LWFA\cite{Malka:2012bi}. Passive injection schemes like self-injection\cite{Corde:2013gj} and ionization-induced injection\cite{Mirzaie:2015js} 
remain the most popular, due to their simplicity to implement. For self-injection, electron spectra with several peaks have occasionally been observed in experiments, see e.g. Ref.\cite{Walker:2013fc}, yet these are typically the result of injection into different wakefield periods\cite{Lundh:2013kua} and are neither stable nor tunable.
To generate tunable twin-beams it is necessary to trigger the electron injection process at controlled positions, which is hardly possible using these methods. One solution could be the use of dual-color laser pulses\cite{Zeng:2015dv}, yet this technique still awaits demonstration.
Instead, experimentally more challenging active electron injection techniques are required, among which optical injection\cite{Faure:2006vy} and shock-front injection\cite{Schmid:2010ih,Buck:2013gs} are arguably the most promising methods. The former adds a second collider pulse to the setup to create favorable trapping conditions during the interaction with the drive pulse\cite{Faure:2006vy}. While injection may be caused by several different mechanisms\cite{Fubiani:2004vm,Davoine:2009vw,Lehe:2013it}, depending on the exact conditions, all variants of optical injection produce ultrashort, quasi-monoenergetic electron bunches, whose energy can be adjusted moving the collision point along the plasma channel. Shock-front injection is an injection method purely based on plasma density tailoring that uses a sharp blade or wafer to create a supersonic shock in the gas flow of the target. At this shock front, a rapid transition from high plasma density ($n_1$) to low density ($n_0$) occurs, which leads to a sudden increase of the plasma wavelength $\lambda_p$. Accordingly, the laser wakefield expands, which temporarily facilitates trapping of electrons. As for colliding pulse injection, the injection position is tunable; in this case by moving the blade with respect to the gas jet. 

\begin{figure}[t]\centering
\includegraphics[width=.9\linewidth]{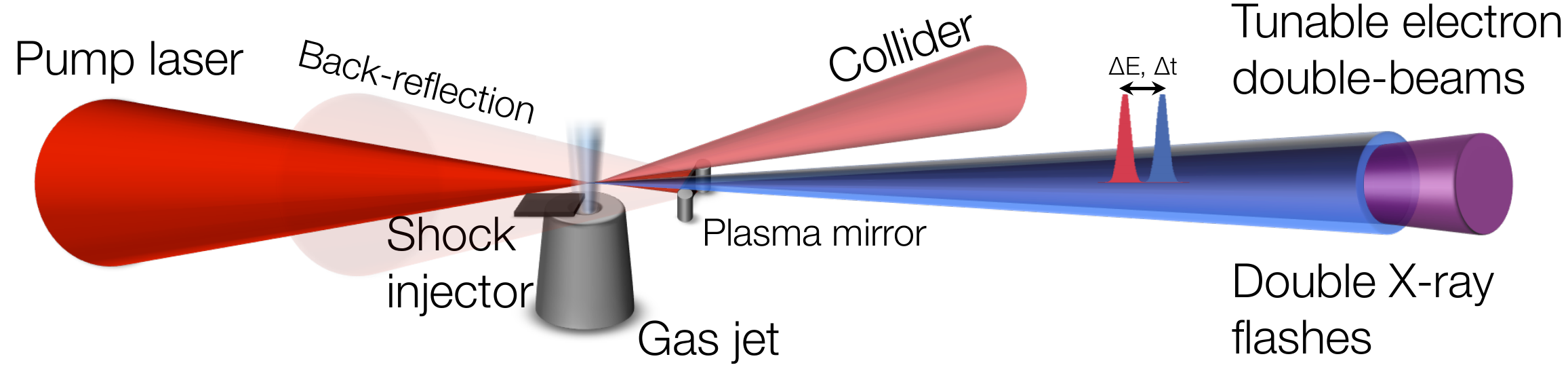}
\caption{Dual-energy femtosecond electron and X-ray source. Using a single high-power laser system, two injection events are triggered by shock-front and colliding pulse injection, resulting in the generation of spectrally distinct electron beams with femtosecond delay. In a second experiment, these electrons are used for radiation generation using Compton backscattering following the reflection of the pump laser on a plasma mirror (Mylar tape).}
\label{fig1_new}
\end{figure}

\begin{figure}[b]\centering
\includegraphics[width=1.\linewidth]{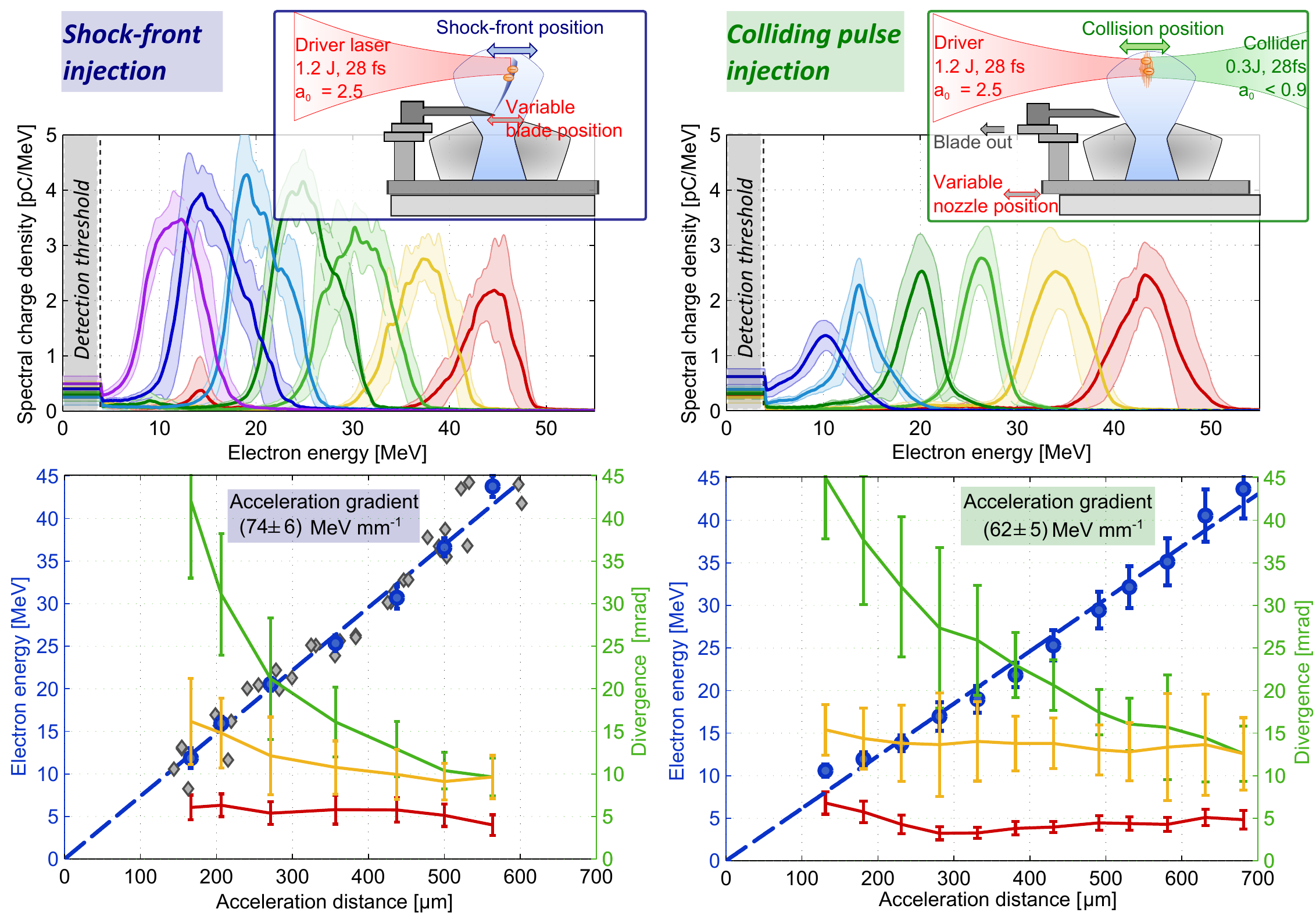}
\caption{Experimental tuning of electron beam energy using shock-front (left) and colliding pulse injection schemes (right). Inset: Sketch of the basic setup. Tuning of the electron energy is obtained by variation of the acceleration length by adjusting the position of the blade (left) or collision position (right) within the gas flow. \textit{Top:} Averaged energy spectra of selected 10 shots for different acceleration length settings. \textit{Bottom:} Resulting beam characteristics. Blue dots represent the averaged peak energy of the bunch, with errorbars showing the rms stability over the complete scan. Red curves depict the average spectral widths (FWHM) for each energy setting. Green (yellow) curves show the (normalized by $\gamma^{3/4}$, see main text) beam divergence. Energy spectra of individual shots of the whole experimental run can be found in the supplementary material.}
\label{fig1_new}
\end{figure}

\begin{figure*}[pt]\centering
\includegraphics[width=1\linewidth]{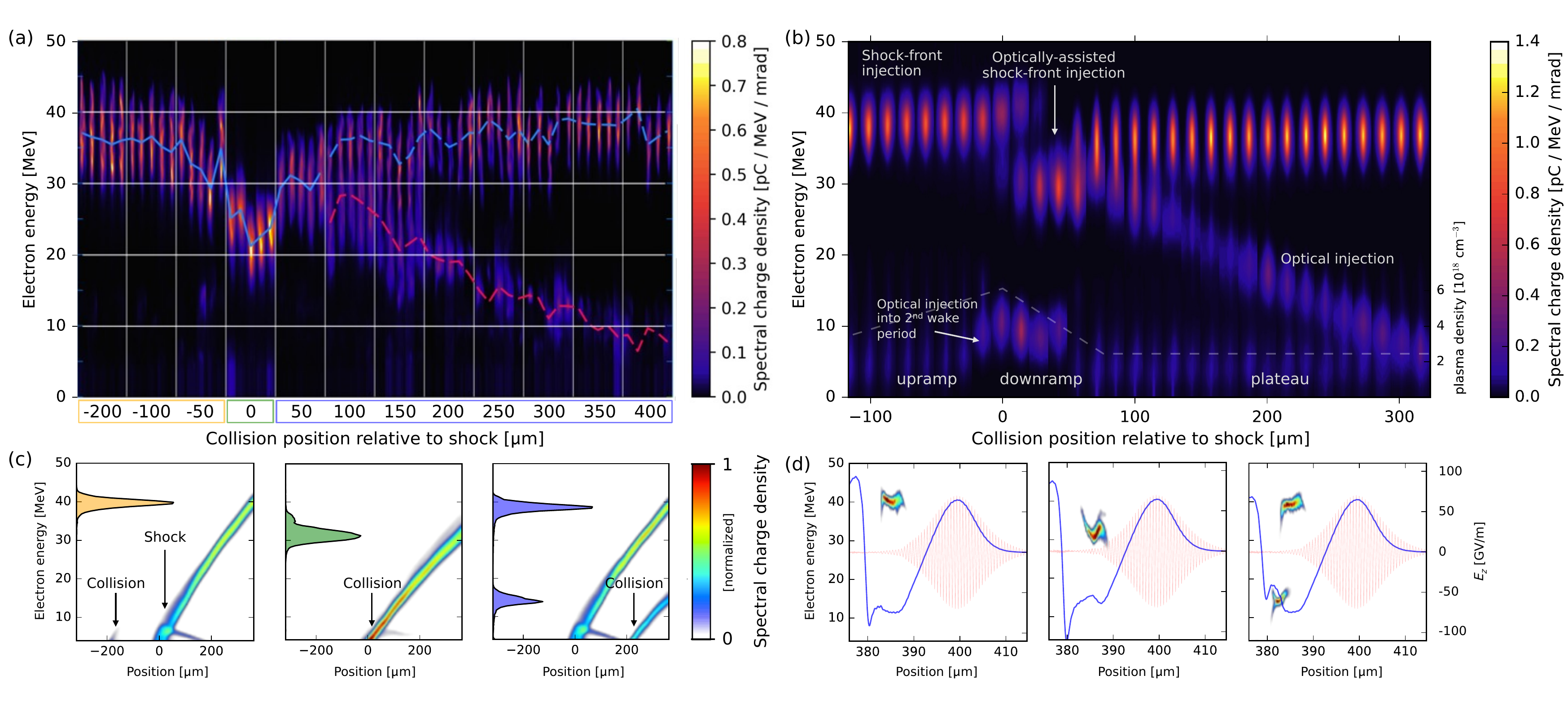}
\caption{Demonstration of dual-energy electron beams and particle-in-cell simulations. (a) Individual electron spectra for different settings of the optical collision position and fixed shock front. Positions are referenced to the shock front. For negative collision position settings, i.e., in the density upramp, no optically injected electrons are observed. Collision at 0 mm shows enhanced injection, while the peak energy (dashed blue line) is reduced due to beam loading. At positive values: Demonstration of the tunable optically injected beam behind the leading electron beam from the shock-front injection scheme (peak energy shown as red line). (b) PIC simulations of the experiment, clearly showing the same three distinct regimes for collision in the upramp, downramp and on the plateau. (c) Evolution of the electron energy spectrum in the first bubble for the three cases along the jet. The final spectrum is indicated along the left vertical axis, with filling colors corresponding to the individual cases (frames of the relative position labels) in (a). The evolution of the spectrum during acceleration is plotted as a 2-D false color plot in each figure. For a collision in the upramp electrons are initially accelerated, but are quickly lost due to the super-luminal phase velocity of the wake. If the collision occurs during the downramp, the amount of injected charge increases, and for collision on the plateau two separate beams emerge. (d) Longitudinal phase space (false color), on-axis plasma field lineouts (blue) and laser field (red) at the end of the accelerator for each case.}
\label{fig2_new}
\end{figure*}


We have compared both techniques in experiments with the ATLAS Ti:Sa laser system, using de Laval gas nozzles to create a supersonic gas flow suitable for shock-front injection (cf. Methods and Figure 1). Figure 2 shows the spectra of electron beams obtained using both shock-front and colliding pulse injection techniques separately, albeit with the same driving laser, focusing geometry, gas nozzle and gas density. The beams exhibit similar properties and have, for an LWFA, a high stability, reproducibility and tuning range. The use of the shock-front injector results in a higher spectral charge density compared to the optical scheme, which is, however, paired with a slightly increased energy bandwidth. Single-shot energy spectra are shown in Supplemental Figure S1.

The beam energy scales approximately linearly with the acceleration distance. This is expected as the acceleration length $L_{acc}< \SI{0.7}{\milli\meter}$ is much smaller than the dephasing length of $L_d\simeq \SI{5}{\milli\meter}$, which we estimate using the scalings from Lu et al. \cite{Lu:2007eb}. We find that the average acceleration field experienced by shock-injected beams is $\SI[parse-numbers=false]{(74\pm6)}{\giga\volt\per\meter}$, while we measure an acceleration gradient of $\SI[parse-numbers=false]{(62\pm5)}{\giga\volt\per\meter}$ for optically injected beams. The difference between gradient measurements seems to be caused by beam loading effects, because the beam charge is not constant over the scan and electron bunches do not necessarily have the same charge density. The beam loading effects are manifested by the diminishing bunch charge and hence higher energy when the blade is close to the gas jet entrance (where $n_1/n_0$ is smaller in the case of shock-front injection). In contrast, the beam charge for colliding pulse injection increases towards the beginning of the jet, which is expected to lower the final beam energy. Here, the bunch charge is related to the product of the vector potentials of collider and main beam\cite{Rechatin:2009uc}, so the increasing charge is likely a result of the relative focusing between the two lasers. For both measurements the divergence of the electrons beams scales as theoretically expected with $\gamma^{-3/4}$, where $\gamma$ is the relativistic factor of the electron beam\cite{Thomas:2010ul}.

\section*{Generation of dual-energy electron beams} 

In a next step, we integrated both injection configurations into the same setup, which allowed us to use them simultaneously. As moving the shock front position perturbs the beginning of the gas jet, we moved the blade to a fixed position which continuously produced electron beams with a peak energy of about \SI{40}{\mega\electronvolt}. Then, the colliding pulse is activated and the entire jet (including the blade which forms the shock) is moved along the laser axis with respect to the collision position. It should be noted that this also changes the intensity of the laser at the shock front, leading to a decrease in the bunch charge as the target is moved closer to the focusing parabola, cf. Supplemental Figure S3.

By changing the relative position of shock and pulse collision, we observe three distinct behaviors, cf. Figure 3a. When the collision occurs before the shock, the electron beam spectra show only a single monoenergetic peak. This signal is indistinguishable from sole shock-front injection, indicating that the optical injection mechanism is suppressed. With the collision occurring close to the density transition, we observe that the energy of the monoenergetic feature suddenly decreases. As this effect is accompanied by an increase in beam charge, it resembles the situation studied by Fubiani et al. \cite{Fubiani:2006}, where colliding pulse injection is augmented by a density down-ramp, and the enhanced charge causes an energy reduction by beam-loading. Then, as we move the collision point further outwards onto the density plateau, two separate monoenergetic peaks form. The first is very similar to shock-front injection without the collision pulse, and is therefore still attributed to shock-front injection. The second, low energy peak could be adjusted between \SI[parse-numbers=false]{10-25}{\mega\electronvolt} by varying the collision position and is hence the result of colliding pulse injection. We also notice that both charge and energy of the optically injected electrons are lower than for the reference case without a shock (cf. Fig.S2). As we discuss below, this effect is attributed to beam loading caused by the shock-injected bunch, which weakens the accelerating fields during injection and subsequent acceleration.


To understand the physics underlying the different operation regimes of the accelerator, we have performed a series of quasi-3D Particle-in-cell simulations. The results, which are depicted in Figure 3b-d, accurately reproduce the experimentally observed behavior. Before the shock, the plasma density is increasing and hence, the plasma wavelength decreases. Analogous to the plasma wave expansion during shock-front injection, this causes a wake contraction at a velocity $v_{ramp}\propto d\lambda_p/dt$. Electron injection and acceleration only occurs if the electrons can reach a velocity close to the wake's phase velocity $v_{\phi}\simeq v_g - v_{ramp}$, where $v_g$ is the group velocity of the laser. During the upramp ($v_{ramp}<0$), the phase velocity is increased and can even reach superluminal speed \cite{Xu:2017ea}. So despite electrons gaining momentum after the pulse collision, injection is inhibited during this phase. The situation changes close to the density peak of the shock, where the relaxed injection conditions allow the pre-accelerated electrons to get trapped inside the second wakefield period. 

Once the pulse collision occurs at the density downramp, the injection behavior changes. A single injection event is observed, with increased charge and lower final beam energy. Simulations suggest that we are operating in the regime of optical transverse injection, where the pulse collision causes a rapid contraction and re-expansion of the wakefield\cite{Lehe:2013it}. Accordingly, this regime can be understood as optically-assisted shock-front injection. As this combined injection occurs inside the same plasma cavity, the increased charge causes beam loading and thus, lowers the energy gain.   

Moving the collision point towards the end of the jet, the two injection events separate, resulting in two distinct peaks in the spectrum. In the simulations, we observe a FWHM duration of \SI[parse-numbers=false]{12-14}{\femto\second} for the high energy bunch, while the duration of the second beam varies from \SI[parse-numbers=false]{4 - 12}{\femto\second}. Note that this is a higher value than typically observed in experiments \cite{Lundh:2011js, Heigoldt:2015cd,Xu:2017ea}, but similar to other simulation results e.g. on colliding pulse injection \cite{Lundh:2011js}. The varying duration of the second electron bunch can be explained as a result of beam loading of the first beam, which prohibits injection in its vicinity. This can also be seen in the shape of the simulated longitudinal wakefields (cf. Fig.3d). Accordingly, the beam charge and energy of the optically injected beam are reduced compared to pure colliding pulse injection for injection very close to the shock-injected bunch.

The temporal delay $\Delta t$ between the peaks of both femtosecond electron bunches is determined by two factors, the duration of the first beam and the amount of dephasing that occurs between the injection events. In simulations we observe a robust temporal synchronization, which approximately follows $\Delta t \simeq \tau_{shock}+(1-v_g/c_0)\cdot\Delta x/c_0$, where $\tau_{shock}$ is the duration of the shock-injected beam and $\Delta x$ is the distance between shock and pulse collision (see supplemental material for more details). In practical numbers this means that $\Delta t$ increases with the injection position by $\SI{3}{\femto\second}\cdot n_e[\SI[parse-numbers=false]{10^{18}}{\per\cubic\centi\meter}]\cdot\Delta x[\si{\milli\meter}].$ Accordingly, the timing jitter between both beams is also very small, as any initial jitter $\Delta t_{jitter}$ between main pulse and collider results in a much smaller timing difference of approximately $\left( 1 - v_g/c_0 \right)\Delta t_{jitter}.$

\begin{figure*}[t]\centering
\includegraphics[width=0.8\linewidth]{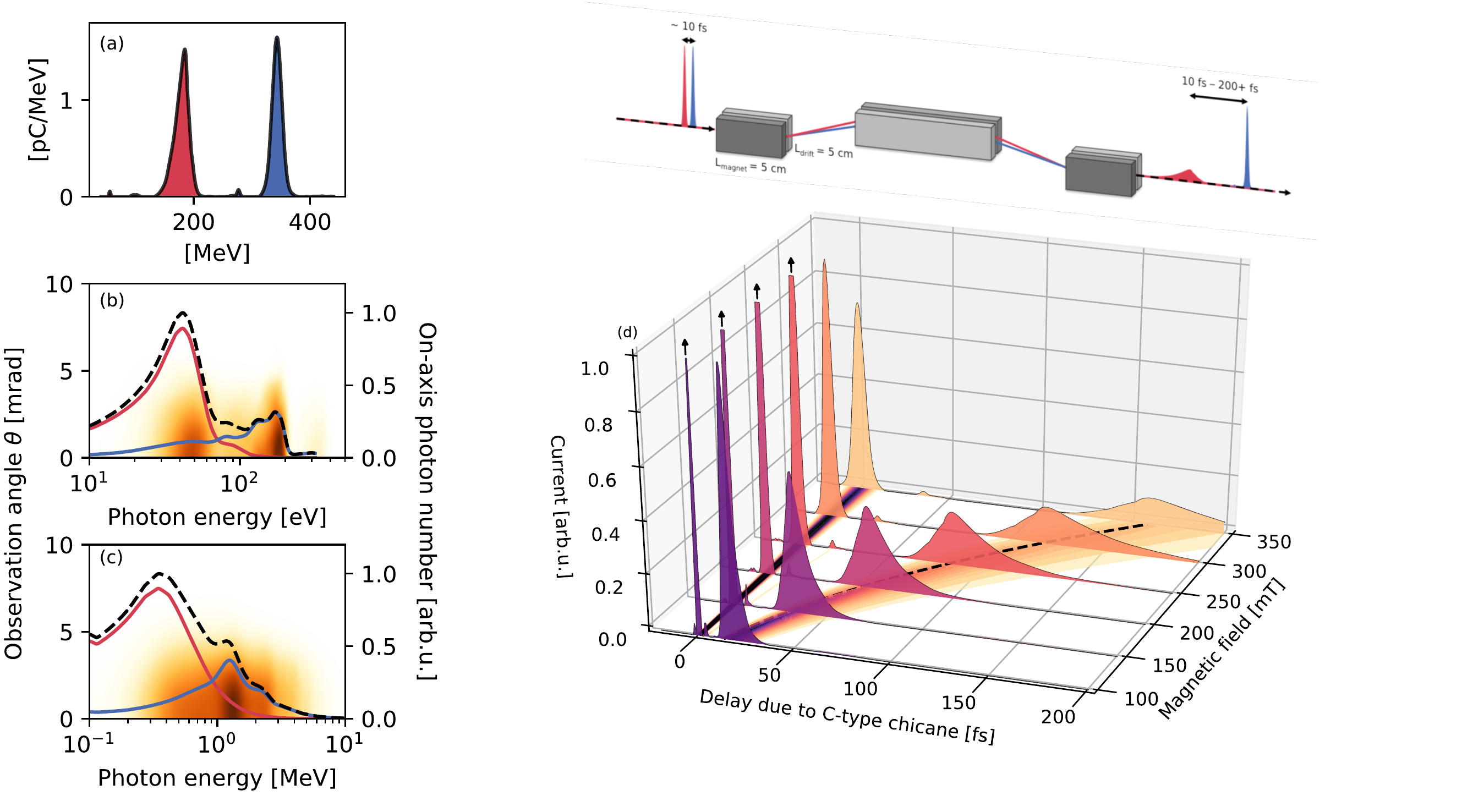}
\caption{Proposed radiation source and timing control based on dual-energy electron beams. (a) Electron beam spectrum obtained using the upgraded ATLAS-300 laser, exhibiting two peaks at \SI{185}{\mega\electronvolt} (red) and \SI{343}{\mega\electronvolt} (blue), with an FWHM energy spread of \SI{13}{\percent} and \SI{6}{\percent}, respectively. Plots (b) and (c) show the simulated synchrotron emission in a magnetic undulator ($\lambda_0=\SI{5}{\milli\meter}$, $K=0.55$ and $n_{periods}=60$) and an all-optical Compton source ($\lambda_0=\SI{800}{\nano\meter}$, $a_0=1.5$, $\Delta t = \SI{30}{\femto\second}$), respectively. Dashed lines show the combined on-axis photon number whereas red and blue lines follow the signal emitted by the respective electron bunches. The underlying colormap shows the angularly resolved radiation emission. Plot (d) shows the delay the electron beams would experience in a C-type magnetic chicance for various magnetic field strengths. The initial bunch duration and delay are estimated to amount to $\lesssim\SI{10}{\femto\second}$.}
\label{fig2}
\end{figure*}

 \begin{figure*}[t]\centering
\includegraphics[width=.65\linewidth]{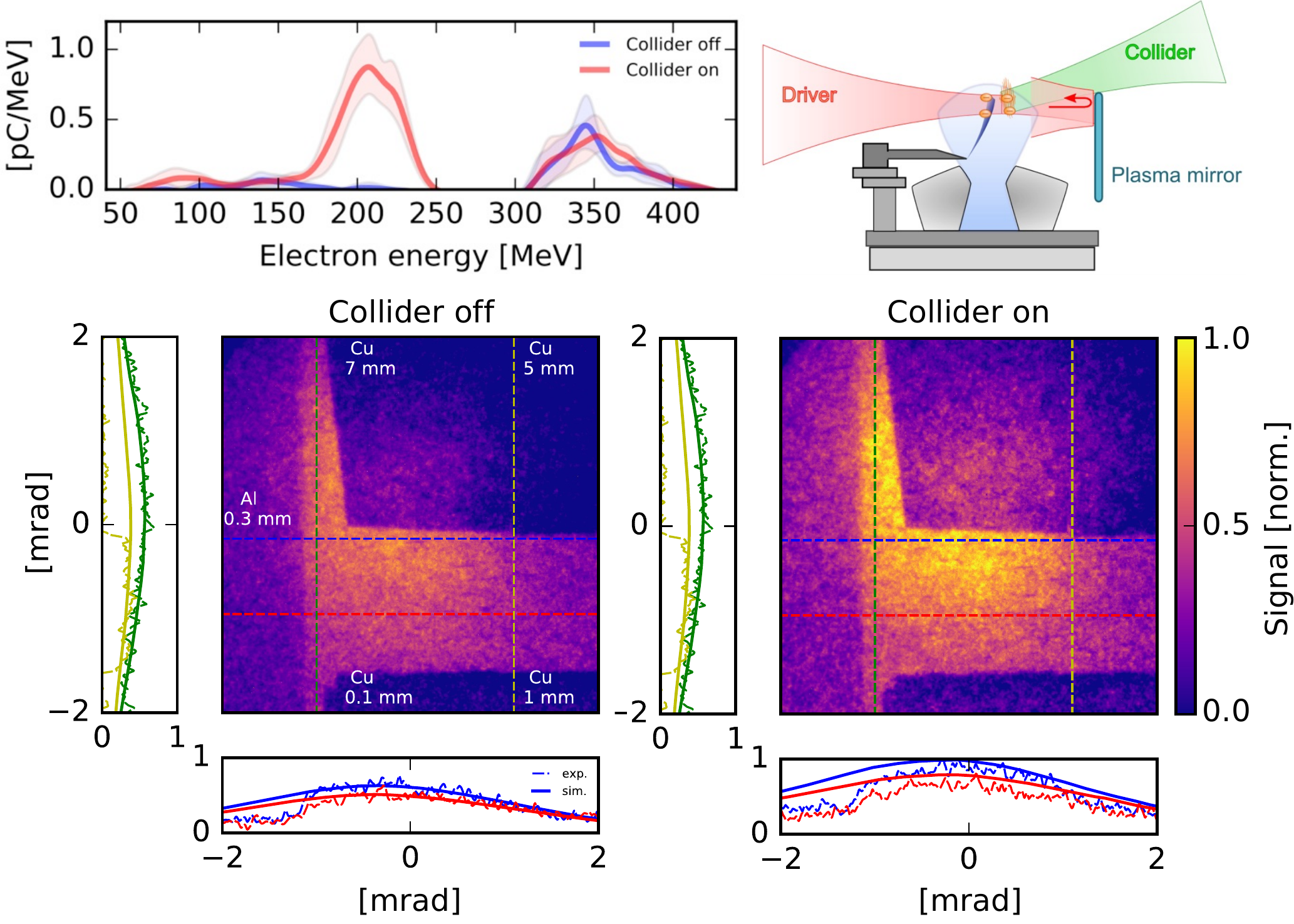}

\caption{Radiation generation using an all-optical Compton source. \textit{Top right: } Sketch of the setup, consisting of a shock-front injector, colliding pulse injector and plasma mirror. \textit{Top left: }Average electron beam spectrum over 10 shots. \textit{Bottom: }Observed radiation signal from Compton backscattering (10 shot average) for a single-energy electron beam (\textit{left}) and enhanced signal using a dual-energy beam (\textit{right}). Both images are normalized to the peak signal level of the dual-beam case, at identical camera settings. Note that darker areas are covered with transmission filters (copper or aluminum, see labels in the plot). Lineouts show that both divergence and peak intensity of the measured data (dashed, light lines) agree within the filter-free areas with the simulated intensity profiles (solid lines) for the same electron beams and a colliding beam with $a_0=1.5$. }
\label{fig5_ComptonExp}
\end{figure*}

Hence, we have demonstrated dual-energy beam generation in an LWFA and simulations accurately reproduce the behavior observed in experiments. Our analysis shows that the electron beam not only exhibits distinctive energy peaks, but also consists of two femtosecond electron bunches with femtosecond delay. Simulations strongly suggest that this delay can be easily modified on a femtosecond scale by changing the distance between injection events, which still needs to be verified experimentally in future. Hence, the results are extremely promising for a variety of applications in ultrafast science. For instance, they could allow the use of femtosecond electron diffraction of the second bunch \cite{He:2016bd} to study femtosecond radiolysis induced by the first\cite{Gauduel:2010ck}. Laboratory-scale dual-color photon sources derived from the electron beams may pose even greater interest, as we will discuss in the following.


\section*{Applicability as secondary radiation source}

 As mentioned earlier, LWFAs can be combined with a number of radiation mechanisms to generate femtosecond photon beams. In the case of a dual-energy electron beam, each electron bunch will emit radiation at different wavelength. For magnetic undulators or Thomson sources, the fundamental wavelength of the emitted photons will be $\lambda_{i}'=\frac{\lambda_0}{\kappa \gamma_i^2}\left(1+\frac{1}{2K^2}\right)$, where $\gamma_i$ is the Lorentz factor of the bunches. In a magnetic undulator the fundamental wavelength $\lambda_0\sim \si{\centi\meter}$, $\kappa=2$ and the undulator strength parameter $K=\frac{e B \lambda_0}{2 \pi m_e c}$, where $B$ is the magnetic field. For Thomson sources, also frequently referred to as inverse Compton sources, the peak angular deflection is determined by the normalized vector potential $a_0$ instead of $K$. Furthermore, the Doppler up-shift depends on the collision angle $\phi$, i.e. $\kappa\simeq 2(1-\cos\phi)$. Due to the small oscillation length $\lambda_0\sim \si{\micro\metre}$, inverse Compton sources can provide keV-scale radiation using $>10$ MeV electrons,\cite{Khrennikov:2015gxa} whereas a comparable undulator source would require GeV-scale electron energies. A Compton source based on the presented dual-energy electron beams would therefore emit dual-color X-rays in the multi-keV regime\cite{Petrillo:2014fo}. As proposed by Kalmykov et al.\cite{Kalmykov:2016kw,Kalmykov:2018ki}, a Compton source combined with a laser-accelerated GeV-scale electron twin-beam could even provide $\gamma$-ray twin-pulses.

Operating an LWFA at higher energy generally reduces the electron beam divergence and gives access to a larger range of useful secondary sources. We have therefore repeated the experiment with an upgraded laser and longer gas jets, which allowed us to increase the electron energy by almost an order of magnitude. Electrons from shock-front injection routinely reached 350 MeV, while the energy of the optically injected beam was tuned from \SIrange{50}{200}{\mega\electronvolt}. Fig.\ref{fig2}a shows an example of a dual-energy beam generated in this new experiment, which has a two quasi-monochromatic peaks at \SI{185}{\mega\electronvolt} and \SI{343}{\mega\electronvolt}. Furthermore, the beam divergence is reduced to \SI[parse-numbers=false]{\sim 1}{\milli\radian}. Simulations show that these electron beams would be suitable to generate XUV radiation (using magnetic undulators, Fig.\ref{fig2}b) or even $\gamma$-rays (based on Compton backscattering, Fig.\ref{fig2}c). Importantly, the temporal profile of the backscattered radiation will be determined by the electron beam\cite{Corde:2013bja}. For weak undulator parameters ($K<1$) the radiation from both electron bunches is spectrally distinguishable, while the emission from non-linear Compton scattering is more broadband and partly overlaps. The latter could be improved using chirped scattering pulses \cite{Rykovanov:2016dv}. 
As discussed previously, simulations predict that the laser-accelerated dual-energy electron beams exhibit a clear temporal structure, with the high energy bunch arriving first, followed by the second femtosecond electron bunch \SI[parse-numbers=false]{10-20}{\femto\second} later. The radiation emitted by these electrons will therefore also take the form of two femtosecond bursts.

As a proof-of-principle experiment, we have performed inverse Compton scattering with both single and dual-energy electron beams, see Fig.\ref{fig5_ComptonExp}. For this, we integrated a plasma mirror into the experimental setup that retro-reflects the drive laser pulse of the LWFA onto the trailing electrons\cite{TaPhuoc:2012cg}. The upper edge of the plasma mirror was placed close to the laser axis, allowing the colliding beam to pass over it, cf. Fig.\ref{fig5_ComptonExp}. In this experiment, the 350-MeV-beam from shock-front injection had a comparably low charge of \SI[parse-numbers=false]{(25\pm 10)}{\pico\coulomb} and we observed a \SI{2.3}{\milli\radian} FWHM photon beam on the detector. Activating the collision pulse added a quasi-monochromatic bunch at \SI{210}{\mega\electronvolt} with \SI[parse-numbers=false]{(39\pm 10)}{\pico\coulomb} charge to the electron beam, while the shock-injected beam remained unchanged with a charge of \SI[parse-numbers=false]{(27\pm 9)}{\pico\coulomb}. Under these conditions the peak signal on the scintillator was \SI{45}{\percent} higher and the beam divergence increased to \SI{2.8}{\milli\radian}, showing clearly the contribution of the second electron bunch. Simulations based on the measured electron beams predict a similar signal gain (\SI{57}{\percent}) and reproduce the beam profile (see lineouts in Fig.\ref{fig5_ComptonExp}). The expected radiation spectrum is similar to Fig.\ref{fig2}c.

Therefore, the presented dual-energy source has great potential to provide dual-color X-rays. However, before these beams can reliably be used for such applications, issues such as beam transport need to be considered. In particular, the transport of beams with large energy separation will require higher-order magnetic optics or beam optics with low chromaticity, see also supplemental material.


The delay between both electron bunches, and hence all derived radiation sources, is determined by the injection mechanism. In our measurements we reproducibly observe the emission of injection radiation\cite{Thomas:2007cc} at the shock and collision positions. As discussed above, this will result in essentially jitter-free delay between both beams, which are accelerated within the same ion cavity. But for many applications, such as already mentioned pump-probe measurements or also plasma wakefield acceleration\cite{Blumenfeld:2007ja}, it is also important to be able to fine-tune the delay of the electron twin-bunches. Given the large energy difference achieved with this source, the bunch separation naturally increases over a propagation distance $x$ due to velocity delay by $\Delta t \simeq \left( \frac{1}{\gamma_1^2} -  \frac{1}{\gamma_2^2} \right)\frac{x}{2}$. For the electron beam from Fig.\ref{fig2}a this evaluates to $\Delta t \simeq \SI{9}{\femto\second}\times x[\si{\meter}]$. However, at the same time the entire beam will also experience temporal broadening due to its divergence. While the latter dominates for beams of large divergence, we concentrate in the following on \SI{1}{\milli\radian} electron beams, as generated in the last experiment. Here the divergence-induced pulse broadening remains on the order of \SI[parse-numbers=false]{1-2}{\femto\second} for meter-scale propagation lengths and is therefore negligible.

Similar to twin-beam approaches at FELs, even longer delays can be achieved using magnetic chicanes. Again, the large energy difference is beneficial, because it allows to adjust the electron bunch delay over short distances. As shown in Fig.\ref{fig2}d, a 30-cm-long chicane with modest magnetic fields (\SI[parse-numbers=false]{100-350}{\milli\tesla}) would be sufficient to induce a delay ranging from a few fs to more than \SI{200}{\femto\second}, which is the timing range required for pump-probe measurements of atomic and molecular systems. In this regime the initial delay and duration of the electron beams ($\sim\SI{10}{\femto\second}$) is also negligible. Nonetheless, the exact duration and delay would have to be determined in future experiments in order to increase the resolution from $\sim \SI{10}{\femto\second}$ to about $\SI{1}{\femto\second}$. While the finite energy spread of each electron bunch will lead to pulse lengthening (especially for the low energy beam), the bunch length remains much shorter than the delay. This is a critical criterion for applications such as sequential imaging or molecular excitation and probing, see Supplemental Material for more details on potential applications.

Last, it should be mentioned that the high-energy pulse always arrives first in our configuration. This is due to the injection and dephasing process in an LWFA and it would be preserved in the chicane setup, as well. The time-energy correlation can to some extent be compensated within the LWFA itself\cite{Dopp:2018kf}, but a complete inversion will be limited by pump depletion and other related processes that occur at the end of the accelerator. Still, the presented time-energy correlation is suitable for many applications such as the ones mentioned above. 

\section*{Conclusions}

In conclusion, we have proposed and demonstrated a scheme to generate monoenergetic femtosecond electron twin-beams with tunable energy and delay using an LWFA. The setup can be implemented at any 100-TW-scale laser facility and generates co-linear electron beams with largely different energies, while the timing jitter is estimated to be of the order of femtoseconds, which is advantageous compared to e.g. a setup based on two separate LWFAs. The double-bunch structure was successfully produced at maximum energies of \SI{40}{\mega\electronvolt} and  \SI{350}{\mega\electronvolt}, and in a proof-of-principle experiment the system was combined with an all-optical Compton source. 

Our analysis shows the complex interplay of the two different electron injection schemes and we have identified regimes for twin-beam generation as well as a new regime of optically-assisted shock-front injection. Further studies may also allow generating specific longitudinal charge density profiles to optimize the accelerator performance\cite{Manahan:2017ky}, probe the beam loading by the first beam\cite{Rechatin:2009ela} or study plasma wakefield acceleration schemes\cite{Hidding:2010es}. 

Furthermore, we have presented a detailed discussion of the source's potential as secondary radiation source. As all laser-wakefield sources, the electron beams can be combined with various radiation mechanisms to generate short-wavelength photon beams in the ultraviolet to $\gamma$-ray range. In particular, we propose to use the source for dual-color pump-probe experiments. Here our approach has many advantages to existing dual-color sources, i.e. it is essentially jitter-free, the pulse delay can be fine-tuned in the fs-ps range and it can operate at very large energy separation. At the same time, all electron beams and secondary radiation sources are inherently synchronized to the drive laser system. Hence, this laboratory-scale source would be suitable to study the ultrafast behavior of complex molecular systems as required for many problems of ultrafast biology\cite{Sundstrom:2008gp}, chemistry\cite{Zewail:2000co} and physics\cite{Ullrich:2012jh}.

\small{
\section*{Acknowledgements}

This work was supported by DFG through the Cluster of Excellence Munich-Centre for Advanced Photonics (MAP EXC 158), DFG-Project Transregio TR-18 funding schemes, by EURATOM-IPP and the Max-Planck-Society. L. V. acknowledges the support by a grant from the Swedish Research Council (2016-05409). The authors thank F. Krausz for helpful comments. A.D. thanks I. Andriyash (WIS) for support with \textsc{Chimera}.

\section*{Author contributions}
 
A.B., M.H., K.K., J.W., J.X., L.V. and S.K. performed the experiments with ATLAS-60 at the MPQ. A.D., H.D., M.G., J.G., S.S. and S.K. performed the experiments with the upgraded laser system at LEX Photonics. A.D., K.K., S.S. and J.W. analyzed the experimental data. A.D. performed PIC simulations, radiation and beam transport calculations. A.D., W.H., K.K., J.W., L.V. and S.K. discussed the results. A.D., K.K. and J.W. wrote the paper. S.K. supervised the project.

$^*$ These authors contributed equally to this manuscript. 

$^\dagger$ Corresponence should be addressed to Andreas D\"opp (a.doepp@physik.uni-muenchen.de) or Stefan Karsch (stefan.karsch@mpq.mpg.de).

\section*{Data availability}
The data that support the plots within this paper and other findings of this study are available from the corresponding authors upon reasonable request.

\newpage


\section*{Methods}
\textbf{Laser systems.} The first series of experiments demonstrating low-energy electrons ($\SI[parse-numbers=false]{<40}{\mega\electronvolt}$) were performed using the ATLAS Ti:Sapphire laser at the MPI for Quantum Optics (MPQ), which delivered laser pulses with \SI{1.5}{\joule} energy within \SI{28}{\femto\second} duration (\SI{50}{\tera\watt}), centered at \SI{800}{\nano\meter} wavelength. The main part of the laser (\SI{1.2}{\joule}) has been focused with an F/13 off-axis parabolic mirror reaching an intensity of \SI{1.3e19}{\watt\per\square\centi\meter} ($a_0=2.5$) in a focal spot of $\SI[parse-numbers=false]{(11\times 12)}{\square\micro\meter}$, at $\SI[parse-numbers=false]{\sim 1.5}{\milli\meter}$ above the nozzle exit. 
For colliding pulse injection, a small part of the beam containing  $\SI[parse-numbers=false]{\sim 0.3}{\joule}$ has been cut off by a pick-up mirror in the experimental chamber and was focused by an F/26 off-axis parabolic mirror. For perfect beam overlap, this would result in a peak potential of up to $a_1\simeq0.9$. However, the actual intensity during the pulse interaction is expected to be much lower, because during alignment we do not optimize towards highest overlap (as in experiments on Thomson-Backscattering\cite{Khrennikov:2015gxa}), but for best energy spread and stability of the optically-injected electron bunch. 

High-energy experiments have been carried out with the upgraded ATLAS laser system, situated at the Laboratory for Extreme Photonics (LEX) at Ludwig-Maximilians University of Munich (LMU), which delivered up to \SI{2.5}{\joule} pulses of similar duration (\SI[parse-numbers=false]{\sim80}{\tera\watt}). The pulses were focused on the gas target in a f/25 geometry to a peak intensity of \SI{5.5e18}{\watt\per\square\centi\meter}, while the colliding pulse had the same parameters as in the first experiment.

\textbf{Gas targets and diagnostics.} In the first experiments, the gas target was a \SI{300}{\micro\meter} supersonic gas nozzle of de Laval geometry. The shock front was realized by introducing a razor blade perpendicular to the gas flow (see Figure 1). In the second experiment, a larger \SI{5}{\milli\meter} nozzle was used in conjunction with an adjustable silicon wafer. The density profile is characterized using both interferometry and few-cycle shadowgraphy (in the second experiment).
The electron energy was characterized using dipole magnet spectrometers. During the first experiment, a magnet spectrometer consisting of a \SI{0.91}{\tesla} permanent magnet, resolving electron energies from \SIrange{2.5}{400}{\mega\electronvolt}.\cite{Sears:2010bt} In the second experiment, an \SI{80}{\centi\meter} permanent magnet (\SI{0.85}{\tesla}) was used, measuring electron energies from \SI{50}{\mega\electronvolt} onwards. The plasma mirror for all-optical Compton backscattering is generated using a \SI{15}{\micro\meter} thick oxide-coated Mylar tape\cite{Shaw:2016is}. The emitted radiation  was detected using a Gd$_2$O$_2$S:Tb scintillator, which is fiber-coupled to an MCP-based image intensifier and whose amplified signal is coupled to a CCD sensor \cite{Gotzfried:2018cx}. Note that the intensifier leads to a decrease in image resolution.

\textbf{Particle-in-cell simulations.} The simulations were performed with the quasi-3D code \textsc{Calder-Circ}\cite{Lifschitz20091803}, using the two modes m=0 and m=1 for modeling of the laser and wakefield. The resolution chosen for the parameter scan was $\Delta x= 0.25 k_0^{-1} $, $\Delta r= 1.0k_0^{-1}$ and 40 particles per cell ($k_0^{-1}=\lambda_0/2\pi \simeq \SI{127}{\nano\meter}$ ). The laser driver is initialized with a peak potential $a_0 = 2.5$ and a spot size of \SI{12}{\micro\meter}, whereas the collider has an intensity $a_1=0.3$. As the density transition length was not directly measured in the first experiment, a plasma gradient length similar to Swanson et al.\cite{Swanson:2017gn} was chosen ($\SI{75}{\micro\meter}\sim 3.5 \lambda_p$). A total of 40 PIC simulations was performed for the parameter scan shown in Fig.3. We note that some experiments suggest an even shorter transition length\cite{Xu:2017ea}. For this case, simulations still show the same three regimes of operation, but the final energy spread is higher because of accumulated energy chirp at the end of the simulation.

\textbf{Radiation modeling.} Synchrotron radiation was calculated using \textsc{Chimera}\cite{Andriyash:2015th}. For this, the measured electron spectra were modeled using $10^4$ test particles, assuming an FWHM beam divergence of \SI{1}{\milli\radian}. For undulator simulations, the parameters from Fuchs et al.\cite{Fuchs:2009da} were used, i.e. $K = 0.55$, $N=60$ and $\lambda_u=\SI{5}{\milli\meter}$. 
As the exact experimental parameters of the $\lambda_0=\SI{800}{\nano\meter}$ scattering pulse for the all-optical Compton source could not be measured in our setup, we used PIC simulations to estimate the normalized peak potential at the end of the gas target ($a_0\sim 2$). Taking into account the reflectance of the plasma mirror ($\sim 0.5$), we chose $a_0=1.5$ as estimation for the scattering parameter. The duration is assumed as \SI{30}{\femto\second} FWHM. To generate the simulated beam profiles, the average electron spectra of 10 shots each were used to calculate the energy-depended far-field emission. In this case, a divergence of \SI{1.5}{\milli\radian} was used for the low-energy beam and the detector was modeled based on its quantum efficiency.

\textbf{Beam transport model.} The delay in a magnetic chicane presented in Fig.\ref{fig2}d is calculated based on matrix transport elements for ideal dipole magnets and velocity delay in a c-type chicane consisting of four 5-cm-long dipole magnets with varying field strength and two 5-cm-long drifts between the magnets. For the first electron beam (\SI{343}{\mega\electronvolt}) we calculate a linear transport matrix element $R_{56}=\SI[parse-numbers=false]{-(3.2-39)}{\micro\meter}$ and a second order matrix element $T_{566}=\SI[parse-numbers=false]{(4.8-58)}{\micro\meter} $ at \SI[parse-numbers=false]{100-350}{\milli\tesla} field strength. For the second bunch (\SI{185}{\mega\electronvolt}), $R_{56}=\SI[parse-numbers=false]{-(11-134)}{\micro\meter}$ and $T_{566}=\SI[parse-numbers=false]{(16.4-201)}{\micro\meter}$. For the FWHM energy spreads of both beams (\SI{5.8}{\percent} for the high energy bunch and \SI{12.8}{\percent} for the low energy bunch) this corresponds to a bunch lengthening of \SI[parse-numbers=false]{(0.6 - 6.9)}{\femto\second} and \SI[parse-numbers=false]{(3.8 - 46.0)}{\femto\second}, respectively. Electron beam divergence is neglected in these calculations since the divergence induced pulse broadening is negligible (\SI{0.5}{\femto\second} for $L_{chicane}=\SI{30}{\centi\meter}$).

\end{document}